\documentclass[prb,twocolumn,nobibnotes,groupedaddress]{revtex4}
\usepackage{enumerate}
\usepackage{color}
\usepackage{amssymb}
\usepackage{amsmath}
\usepackage{graphicx}
\usepackage{hyperref}
\renewcommand\footnotemark{}

\begin{document}

\newcommand{\tshc}[1]{\textcolor{red}{[#1]}}

\title{Modeling singlet fission on a quantum computer}

\date{\today}
\author{Daniel Claudino,$^1$\footnote{\href{mailto:claudinodc@ornl.gov}{claudinodc@ornl.gov}} Bo Peng,$^{2}$\footnote{\href{mailto:peng398@pnnl.gov}{peng398@pnnl.gov}} Karol Kowalski$^{2}$ and Travis S. Humble$^{3}$}
\affiliation{
$^{1}$Computational Sciences and Engineering Division,\ Oak\ Ridge\ National\ Laboratory,\ Oak\ Ridge,\ TN,\ 37831,\ USA \\
$^{2}$Physical Sciences Division, Pacific Northwest National Laboratory, Richland, WA 99352, USA\\
$^{3}$Quantum Science Center,\ Oak\ Ridge\ National\ Laboratory,\ Oak\ Ridge,\ TN,\ 37831,\ USA}

\thanks{This manuscript has been authored by UT-Battelle, LLC, under Contract No.~DE-AC0500OR22725 with the U.S.~Department of Energy. The United States Government retains and the publisher, by accepting the article for publication, acknowledges that the United States Government retains a non-exclusive, paid-up, irrevocable, world-wide license to publish or reproduce the published form of this manuscript, or allow others to do so, for the United States Government purposes. The Department of Energy will provide public access to these results of federally sponsored research in accordance with the DOE Public Access Plan.}
\begin{abstract}

We present a use case of practical utility of quantum computing by employing a quantum computer in the investigation of the linear H$_4$ molecule as a simple model to comply with the requirements of singlet fission. We leverage a series of independent strategies to bring down the overall cost of the quantum computations, namely 1) tapering off qubits in order to reduce the size of the relevant Hilbert space; 2) measurement optimization via rotations to eigenbases shared by groups of qubit-wise commuting (QWC) Pauli strings; 3) parallel execution of multiple state preparation + measurement operations, implementing quantum circuits onto all 20 qubits available in the Quantinuum H1-1 quantum hardware. We report results that satisfy the energetic prerequisites of singlet fission and which are in excellent agreement with the exact transition energies (for the chosen one-particle basis), and much superior to classical methods deemed computationally tractable for singlet fission candidates.

\end{abstract}


\maketitle

\section{Introduction}
\label{sec:intro}

The research, engineering, and manufacturing of materials with specific energetic capabilities are growing unquestionably more important as a steadily increasing demand motivates the adoption of more efficient methods for energy generation, storage, and transport. This spurs research in many areas, with scientific and technological developments in chemistry and materials science taking the center stage. These efforts largely benefit from quantum mechanical models that describe and predict matter at a fundamental level, the phenomenon of singlet fission being a primary example.\cite{singlet_fission1, singlet_fission2} In short, knowledge of the electronic/excitonic structure of molecular systems enables the search and engineering of species where the first excited singlet state ($S_1$) sits twice as high (usually slightly higher) in the energy scale as the lowest energy triplet state ($T_0$). In a pair of chromophores initially at the ground state $S_0$, one is excited to the $S_1$ state, which evolves by means of internal conversion to two triplet states $T_0$ coupled as a spin singlet composite system.\cite{singlet_fission2, zimmerman2010singlet}


A deeper understanding of the singlet fission phenomenon has been aided by adopting computational tools 
to screen potential singlet fission candidates and probe their energy levels. Density functional theory (DFT) has been used extensively for computational research of materials and carried out with different flavors due to the relatively low computational cost. However, it is well-documented that these approaches are inadequate for the investigation of singlet fission. Although comprised of a series of more elementary steps, singlet fission is overall governed by multi-excitonic processes that are not yet completely understood. This makes time-dependent DFT, a routine formulation of DFT that enables the treatment of excited states, and other theories restricted to transitions dominated by single excitations unsuitable for the current case, as they fail to capture the multi-excitonic nature of the $S_1$ state. On the other hand, methods that provide the appropriate theoretical support by accounting for higher manifolds of excitations are computationally expensive, which often prevents their applicability in the context of singlet fission unless further approximations are imposed. As discussed in Section \ref{ssec:comparison}, even methods deemed ``high-level'' can fall short of the stringency necessary to describe the $S_1$ state.

Some of the recurring difficulties in dealing with the challenge of electronic structure, potentially including identifying candidates for singlet fission, are expected to be alleviated or even removed with the maturation of quantum computing. Quantum computing has been postulated as the most natural means by which quantum processes, like singlet fission, could be simulated. This is because the quantum mechanical model enabled by a quantum computer has the potential to efficiently represent the strong correlations that arise during multi-excitonic processes. Several recent experimental demonstrations provide empirical evidence to support such aspirations as technologically feasible,\cite{google_supremacy, advantage_photonic, borealis} but current quantum computers are not yet capable of rivaling conventional computers and their associated methods. Therefore, identifying problems instances for which a computational advantage may emerge from using quantum computers is an important part of this developing field as well as the broader goal of discovering sustainable sources of energy and new ways for energy capture and transformation.

With this goal in mind, we report in this paper the energetics underlying singlet fission as a use case of practical quantum computing, made affordable by a multi-pronged strategy that results in the reduction of required number of measurements. Moreover, the algorithm at the heart of this study proves to be a quite appealing strategy not only for treating the ground state, but is also able to qualitatively and quantitatively address intricate electronic/excitonic structures that are known to plague most current methods within reach of conventional computation.

\section{Results}

\subsection{Model}

\begin{figure}[h]
    \centering
    \includegraphics[width=.75\columnwidth]{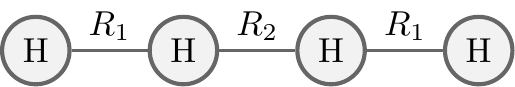}
    \caption{Geometry representation of the linear H$_4$ molecule. This is a valid model for singlet fission for distances $0.5\text{\AA} \leq R_1, R_2 \leq 4.0\text{\AA}$.}
    \label{fig:model}
\end{figure}

In the present study we show how current quantum computers can be employed in studying singlet fission. Given the limited performance window of present quantum computers, the description of $\pi$-conjugated systems, the quintessential species to exhibit singlet fission,\cite{singlet_fission2,singlet_fission3,lewis2016research,ullrich2021unconventional,hong2022steering} is beyond the scope of quantum computation at present. An alternative, much simpler system which has been employed in the modeling of singlet fission is the linear H$_4$ molecule. It has been found that, for specific geometric parameters (Figure \ref{fig:model}), this molecule complies with the requirement that $E(S_0\rightarrow S_1) \approx 2 E(S_0\rightarrow T_0)$. For the sake of simplicity, we choose $R_1=R_2=2${\AA}.\cite{h4_singlet_fission}

The energy levels of the spin singlet and triplet states are determined based on the PDS($K$) energy functional, which belongs to the family of methods grounded in the expansions of the moments of the Hamiltonian and is reviewed in Section \ref{sec:theory}. The PDS formalism enables the efficient re-use of quantum information to calculate multiple states of a quantum system in the course of single simulation providing at the same time a trade-off between quantum circuit depth and the number of measurements. Therefore this formalism can make efficient use of currently existing devices.

\subsection{PDS(\emph{K}) energy functional}
\label{sec:theory}


The PDS($K$) functional refers to the early studies of Peeters and Devreese,\cite{peeters1984upper} later followed by Soldatov,\cite{soldatov1995generalized} in the investigation of the upper bounds for the free energy. We recently reviewed this method and proposed its potential use for accurate quantum computations.\cite{Kowalski2020, claudino2021improving, Peng2021variationalquantum,peng2022State} Briefly speaking, for the equivalent of an expansion of moments of $\hat{H}$ to order $K$, one needs to solve for an auxiliary vector $\mathbf{X} = (X_1,\cdots,X_K)^T$ with $K$ elements from the linear system in Equation \ref{lineq1}
\begin{equation}
\mathbf{MX} = -\mathbf{Y}, \label{lineq1} 
\end{equation}
with the elements of the matrix $\bf M$ and vector $\bf Y$ defined in Equation \ref{eq:elements} as the expectation values of Hamiltonian powers with respect to a reference state $|\phi\rangle$ (i.e. moments), 
\begin{align}
M_{ij} &= \langle \hat{H}^{2K-i-j} \rangle, \nonumber \\
Y_i &= \langle \hat{H}^{2K-i} \rangle,~~ (i,j = 1,\cdots,K) \label{eq:elements}
\end{align}
where, for simplicity we used the notation $\langle \hat{H}^n \rangle \equiv \langle\phi|\hat{H}^n|\phi\rangle$. In the case of a quantum version of the PDS($K$) energy functional, one uses the quantum computer to prepare the trial state $|\phi\rangle$ and perform the measurements necessary to estimate the moments of $\hat{H}$ up to $n=2K-1$. The next step is to feed $\textbf{X}$ as coefficients of the polynomial in Equation \ref{poly}.
\begin{equation}
P_K(\mathcal{E}) = \mathcal{E}^{K} + \sum_{i=1}^K X_i \mathcal{E}^{K-i}=0. \label{poly}   
\end{equation}
The goal of the PDS($K$) calculation is to get the roots of the polynomial, $(a_1^{(K)},\ldots,a_K^{(K)})$. It can be shown that these roots provide upper bounds to the exact ground and excited state energies, e.g., a bound to the ground state energy can be obtained from Equation \ref{eq:gs}
\begin{equation}
    E_0 \le {\rm min}(a_1^{(K)},\ldots,a_K^{(K)}) 
    \le \langle\phi|H|\phi\rangle \;, \label{eq:gs}
\end{equation}
and in like manner, the first excited state is approximated by the second lowest root, and so on.

\subsection{Statevector simulations}

\begin{figure}[ht]
    \centering
    \includegraphics[width=\columnwidth]{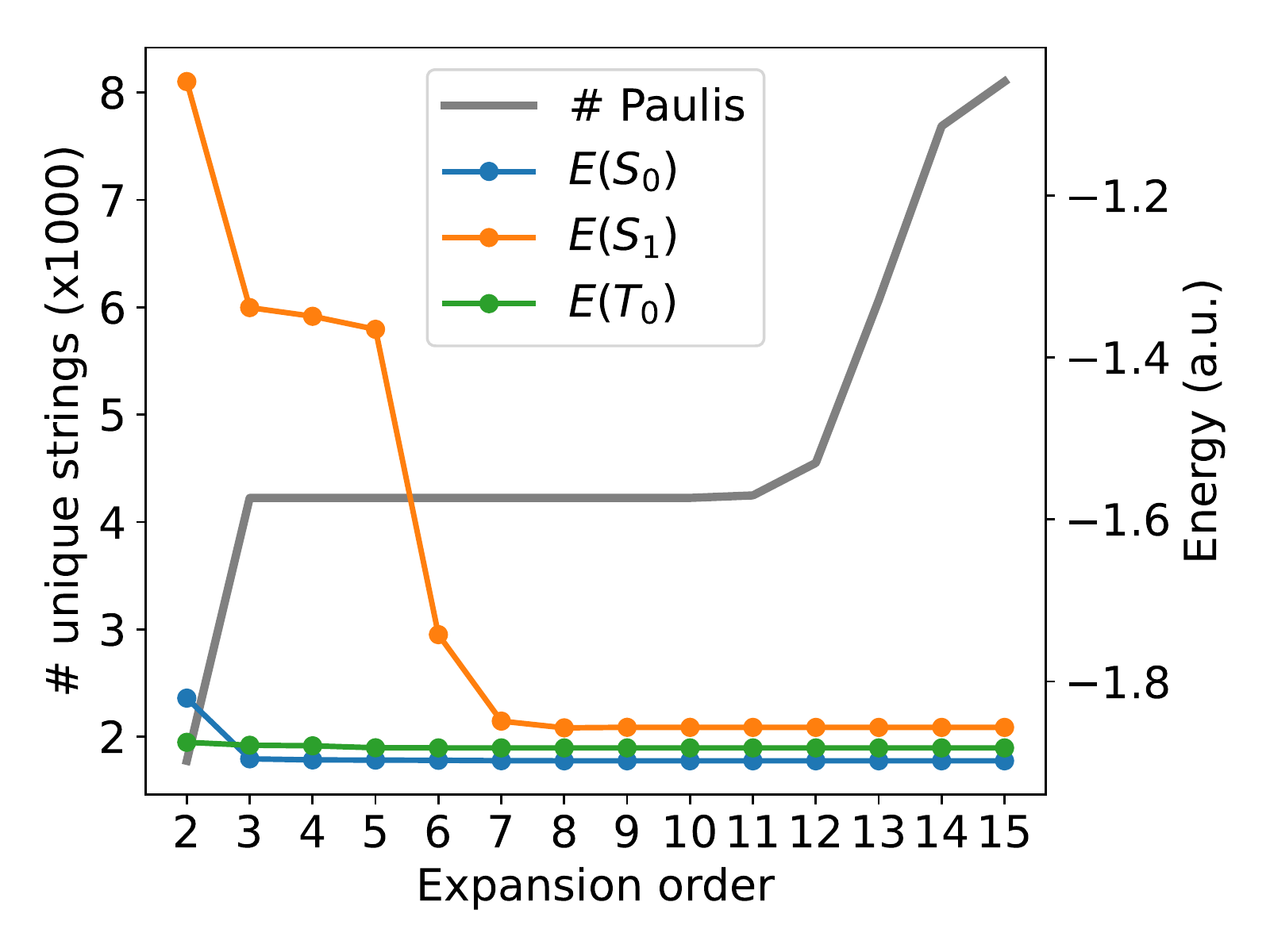}
    \caption{Number of unique Pauli strings and the energies of lowest states of singlet ($S_0$) and triplet ($T_0$) spin multiplicities and the energy of the first singlet excited state ($S_1$).}
    \label{fig:pds_statevector}
\end{figure}

The Horn-Weinstein theorem\cite{horn1984t} states that expansions such as the PDS($K$) energy functional are guaranteed to recover the ground state energy for $K\rightarrow \infty$ if the trial state has non-vanishing overlap with the ground state. Throughout this work we adopted the Hartree-Fock determinant as our trial state, which is a mean-field approximation, and as such, can be prepared without entangling operations. In order to determine a suitable finite order $K$, we run simulations using a noiseless statevector backend (Google's \texttt{qsim}\cite{qsim})
. The energies of the states $S_0$, $S_1$, and $T_0$, together with the number of unique Pauli strings that need to be measured up to $K=15$ are shown in Figure \ref{fig:pds_statevector}.

When computing PDS($K$), one needs to evaluate all moments of the Hamiltonian for $n$, $ 1 \leq n \leq 2K -1 $. It turns out that, as one increases $n$, the Pauli strings that comprise $\langle \hat{H}^{n} \rangle$ can be largely found in the powers $< n$, thus, when estimating a given $\langle \hat{H}^n \rangle$, one needs only to measure strings that are not found in the lower moments. Fortuitously, we see that PDS($K$) has the same measurement cost for $K=3-10$, meaning one can go up to $K=10$ with the same quantum cost as $K=3$, improving upon the results in the course of doing so, while at the marginal classical expense of solving for more roots of the polynomial in Equation \ref{poly}. All three relevant energy values portrayed in Figure \ref{fig:pds_statevector} reach convergence within the $K=3-10$ range, so we will adopt $K=10$ for the remainder of the manuscript.

\subsection{Measurement optimization}
\label{ssec:measurement}

Even when restricting the number of measurements to only the unique Pauli strings, as described in Section \ref{fig:pds_statevector}, in order to compute PDS(10) for both the spin singlet and triplet states it would be necessary to implement a total 8446 unique circuits, each to be run for a given number of shots, which we found to be cost-prohibitive.

A first step into reducing the number of measurements is to taper off a set of qubits onto which the Hamiltonian acts trivially, that is, the outcome is always $\pm 1$ (See Section \ref{ssec:tapering}). Its main purpose is to locate a subspace of the entire 2$^N$-dimensional Hilbert space spanned by $N$-qubit basis states where the ground state can be found, implying the practical advantage that the new Hamiltonian can have at most $2^{N-N_\text{tapered}-1}(2^{N-N_\text{tapered}}+1)$ elements, with $N_\text{tapered}$ being the number of qubits tapered, as opposed to the $2^{N-1}(2^N+1)$ elements in the original Hamiltonian. Even though the Hamiltonian and the moment operators are fairly sparse, that is, the number of non-zero elements is actually much smaller than $2^{N-N_\text{tapered}-1}(2^{N-N_\text{tapered}}+1)$, there is still noticeable reduction in the number of Pauli strings by applying qubit tapering, which yields denser operators in the corresponding subspace.


An alternative avenue that can be employed in conjunction with qubit tapering to achieve further reduction in the necessary resources is to find groups of commuting Pauli strings and to determine a common basis for all strings in the group under some commutativity rule. This change of basis can be realized by a Pauli string that serves as rotation operator to the eigenbasis shared by all Pauli strings in the group, whose expectation values can then be evaluated upon the bitstring counts drawn from measurements on the rotation operator basis. In this paper we adopt qubit-wise commutativity (QWC), meaning a collection of strings qubit-wise commute if the Pauli operators acting in each qubit space commute for all strings in the set, also referred to as finding a tensor product basis.\cite{kandala2017hardware}

The final step in the optimization of quantum resource allocation lies on the recognition that the Quantinuum H1-1 system has a total 20 qubits,\cite{h1} and each circuit under qubit tapering is identified by a 5-qubit string. We therefore leverage a parallelization scheme wherein 4 QWC rotation strings can be measured simultaneously in a single circuit implementation and execution, benefiting from the absence of noise due to crosstalk between qubits in this architecture.\cite{h1_sheet} The resulting number of measurements from the application of the QWC scheme to the original and the tapered Hamiltonians are reported in Table \ref{tab:measurements}, together with the final number of circuit implemented on the H1-1 hardware by exploiting parallelization.

\begin{table}[h]
    \centering
    \caption{Number of circuit implementations in the computation of the PDS(10) energy with the for different measurement optimization strategies.}
\begin{tabular}{c c c}
\toprule
 &  singlet  & triplet \\
 \hline
 Original & 4223 & 4223 \\
 QWC & 441 & 441 \\
 Tapering & 527 & 379 \\
 Tapering + QWC & 122 & 66 \\
  Tapering + QWC + Parallelization & 31 & 17 \\
 \hline
\end{tabular}
    \label{tab:measurements}
\end{table}

\begin{figure*}[!ht]
    \centering
    \includegraphics[width=.9\textwidth]{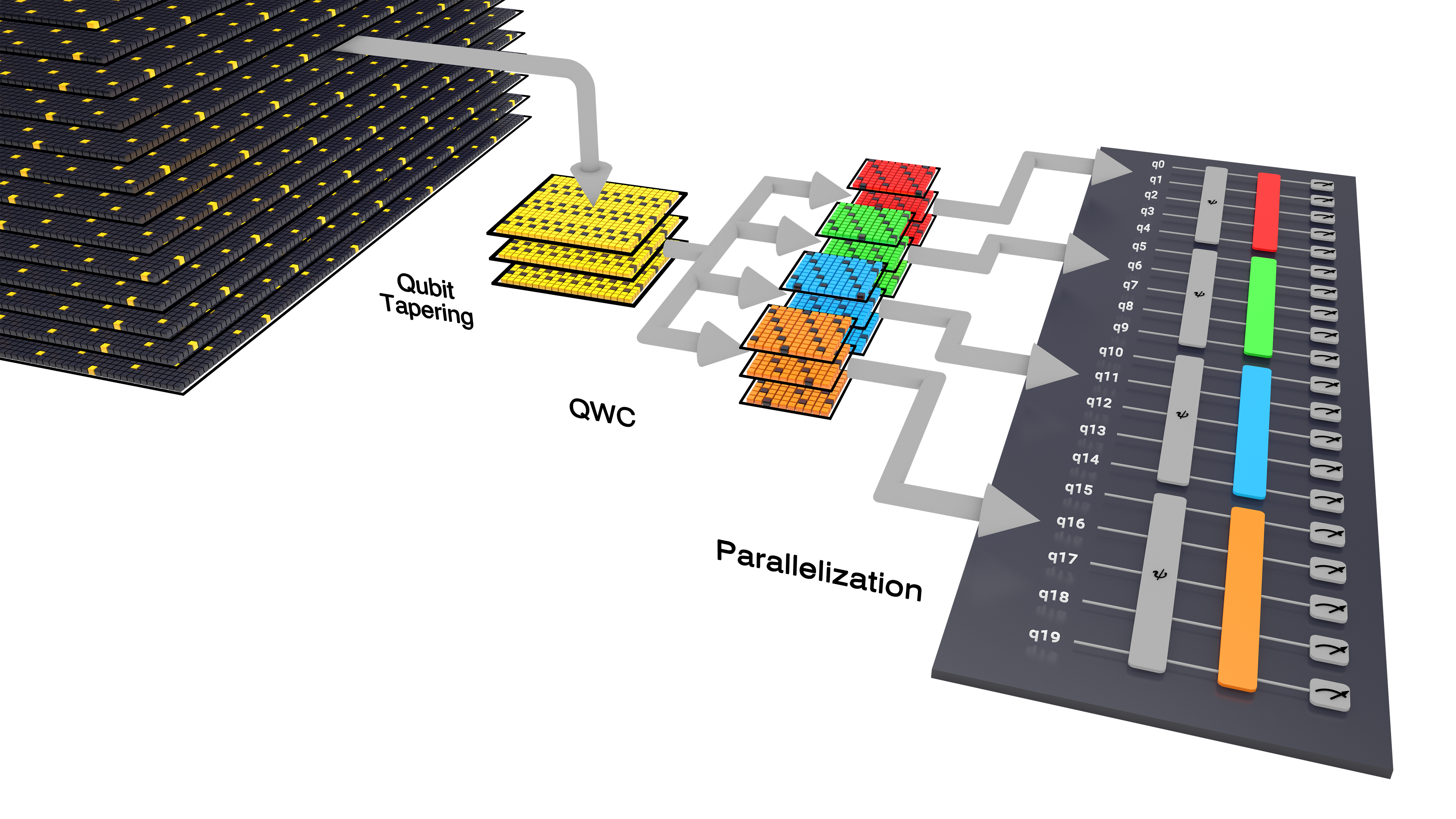}
    \caption{Pictorial representation of the measurement reduction protocol adopted in this paper. From a complex observable comprised of many Pauli strings and whose matrix representation is very large, qubit tapering yieds much smaller and fewer matrices. Those can be arranged in qubit-wise commuting groups (QWC), with four of these being implemented in parallel in the H1-1 quantum hardware.}
    \label{fig:opt}
\end{figure*}

The overall measurement optimization protocol followed here is graphically summarized in Figure \ref{fig:opt}.

\subsection{Quantum simulations and comparison with classical reference methods}
\label{ssec:comparison}

The linear H$_4$ model (with $R-R$ distances set equal to 2\AA) is a commonly used model to study singlet fission processes. This model also epitomizes typical problems quantum chemistry methods stumble into when trying to describe excited states dominated by double excitations. Coupled cluster (CC)\cite{shavitt_bartlett_2009, cc_review} theory is the method of choice for the computation of highly accurate ground state energies, while its equation of motion extension (EOM-CC)\cite{eom, eom1, shavitt_bartlett_2009, cc_review} enables determination of excitation energies with high accuracy, and here we take CCSD, CCSDT, and CCSDTQ and the corresponding EOM counterparts\cite{eomccsd, eom2, eomccsdt1, eomccsdt2} to determine classical reference values, where S, D, T, Q signal that the manifold of excitation operators comprise of all single, double, triple, and quadruple excitations. This series follows a trend of increasing accuracy, while also incurring a pronounced increase in computational cost, to the point that EOM-CCSDTQ involves insurmountable storage requirements for realistic systems. For the sake of notational simplicity, we report classical results by referring to the level of theory employed to calculate the ground state energies, e.g. CCSD, implicitly stating that the energies for the $S_0\rightarrow S_1$ transition were obtained with the EOM variant, EOM-CCSD in this case. Furthermore, because the H$_4$ molecule has four electrons, inclusion of the entire quadruple excitation manifold, as in CCSDTQ, recovers the exact energetics within the one-particle basis in question, being equivalent to diagonalizing the Hamiltonian in the space of all possible particle number-conserving excitations (full configuration interaction), thus we refer to CCSDTQ as ``Exact'' hereafter.

\begin{figure}[!ht]
    \centering
    \includegraphics[width=\columnwidth]{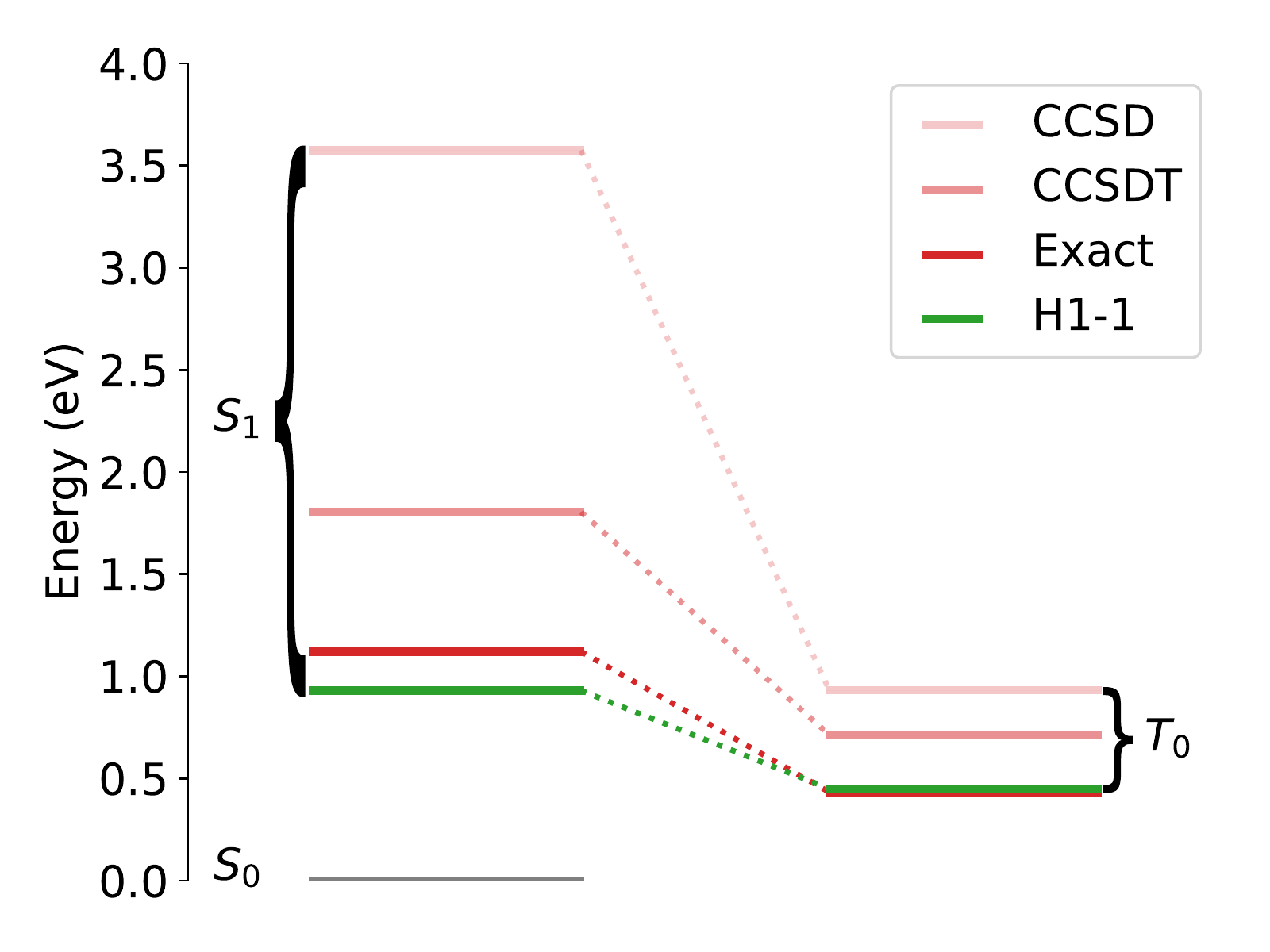}
    \caption{Comparison of excitation energies (in eV) for the lowest singlet excited state 
    ($S_0\rightarrow S_1$) transition and singlet-triplet transition ($S_0 \rightarrow T_0$). The classically computed $S_0\rightarrow S_1$ excitation energies are obtained with EOM variant of the reported CC method.}
    \label{fig:levels}
\end{figure}

With these considerations in mind, one routinely strives for a balance between accuracy and computational cost. However, conventional computing methods for identifying candidates for singlet fission may be characterized by sizable errors in estimated excitation energies for doubly excited states when approximated by a truncation of the excitation manifold. As discussed in the literature,\cite{eomccsdt1, eomccsdt2} these results illustrate typical problems with the EOM-CCSD method in describing more complicated excited states. In this case, one has to incorporate triple or even quadruple excitations in the EOM-CC formalism to arrive at the desired level of accuracy. In Figure \ref{fig:levels} we report the energies gaps involved in the $S_0\rightarrow S_1$ and the $S_0 \rightarrow T_0$ transitions given by PDS(10) from quantum simulations 
on the H1-1 hardware as well as results from CCSD and CCSDT, with the respective EOM transition energies, as well as the corresponding exact results (CCSDTQ). The energy values displayed in Figure \ref{fig:levels}, together with the PDS(10) results from the noiseless statevector simulations are provided in Supplementary Tables I and II.
 
%
%

As seen from Figure \ref{fig:levels}, the CCSD transition energies transition significantly overestimate the exact results, and even the considerably more expensive CCSDT variant falls short of quantitative agreement with the exact energy values. This is a consequence of the doubly-excited character of the $S_1$ state dominated by doubly excited configurations, namely $|\Phi_{2\bar{2}}^{3\bar{3}}\rangle$, $|\Phi_{1\bar{1}}^{3\bar{3}}\rangle$, and $|\Phi_{2\bar{2}}^{4\bar{4}}\rangle$ (see Supplementary Table III). Surprisingly, despite reports in the literature that point to the robustness of CCSDT to effectively incorporate doubly excited configurations, it is clear that this level of theory does not match the demanded accuracy.

Now turning to the results from 
the quantum simulations on the H1-1 hardware, despite the deleterious effect of noise, we notice results that are still superior to the considerably expensive (EOM)CCSDT method. While the $S_0\rightarrow T_0$ achieved by the H1-1 system is virtually identical to the respective exact gap, we observe a larger deviation from the reference energies for the $S_0\rightarrow S_1$ transition. This can be explained bearing in mind that PDS($K$) favors the ground state, and while the $S_0\rightarrow T_0$ transition involves two ground states for different total spin numbers, the $S_0\rightarrow S_1$ transition depends on the performance of PDS($K$) for the $S_1$ state. This is corroborated by Figure \ref{fig:pds_statevector} in how the energies of the states $S_0$ and $T_0$ converge much faster than that of the $S_1$ state. However, it is important to notice that the energy transitions assessed with the H1-1 quantum computer are still in agreement with the requirement that energy gap in the $S_0\rightarrow S_1$ transition is slightly higher in energy than  twice that of $S_0\rightarrow T_0$.  Mitigation of the state preparation and measurement (SPAM) error based on the H1-1 hardware specifications does little to bring the exact and quantum simulated energy gaps closer in the case of the $S_0\rightarrow S_1$ transition, as the source of this discrepancy can be largely attributed to sampling noise. Further details are provided in Supplementary Section III.

\section{Discussion}
\label{sec:conclusion}

One of the biggest challenges posed by the current generation of quantum hardware is the search for computational advantage in problems with real-world implications. With chemistry lending itself as a natural candidate to fulfill this request and invariably being a crucial component of the engineering of specialized materials, the successful application of quantum computation in chemistry carries significant prospects to the area of materials science. In this manuscript we show how quantum computers can be a valuable tool to detail the intricate excitonic structure necessary to enable singlet fission, phenomenon with great interest for the materials with energetic properties.

Several aspects of this work are worth mentioning. We validate the PDS($K)$ energy functional as a viable method to determine the energetics involved in singlet fission, meaning that  itis successfully applied to different spin sectors, and despite clearly favoring the ground state, it is also able to reliably treat demanding excited states which are not satisfactorily captured even by considerably expensive classical techniques. Our multi-pronged approach to measurement optimization leads to a reduction of orders of magnitude in the number of circuit implementations, allowing for simulations that would otherwise be cost-prohibitive. The parallelization of measurements of distinct Pauli strings in the same circuit execution enables the utilization of all 20 qubits available in the H1-1 hardware, without exhibiting major deleterious effect from hardware noise.

\section{Methods}

\subsection{Observables}

We used the XACC quantum computing framework\cite{xacc1, xacc2, xacc3} and its interface with the PySCF quantum chemistry package\cite{pyscf} to obtain the matrix elements to construct the molecular Hamiltonian for the H$_4$ molecule in the STO-3G basis set.\cite{sto3g1} These matrix elements enable the representation of the Hamiltonian in terms of second-quantized fermionic operators, which are in turn mapped onto a sum of tensor products of Pauli spin operators $\hat{P} = \{\hat{X}, \hat{Y}, \hat{Z}, I\}^{\otimes N}$ via the Jordan-Wigner transformation,\cite{JW} where $N=8$ for the original Hamiltonian and $N=5$ after tapering:

\begin{equation}
    \label{eq:h}
    \hat{H}=\sum_ic_i\hat{P}_i,
\end{equation}
where $c_i$ are complex scalars. Similarly, the $n$-th moment

\begin{equation}
    \hat{H}^n=\left(\sum_ic_i\hat{P}_i\right)^n = \sum_ic_i'\hat{P}'_i,
\end{equation}
conforms to the same generic form as in Equation \ref{eq:h}, with primed quantities indicating they differ from those in $\hat{H}$, and $i$ is a dummy index that goes over all scalars/Pauli strings in $\hat{H}^n$. Let us define $\{\hat{P}_{i}\}_k $ the set of Pauli strings in the $k$-th moment. It turns out that, for the $n$-th and $m$-th moments (not necessarily successive)

\begin{equation}
    \{\hat{P}_{i}\}_n  \cap \{\hat{P}_{i}\}_m \neq \{\varnothing\}.
\end{equation}

Therefore, we need only measure the same string once:

\begin{equation}
\label{eq:unique}
    \{\hat{P}_i\}_\text{unique} = \bigcup_{n=1}^{2K-1}  \{ \hat{P}_i\}_n - \bigcap_{n=1}^{2K-1}  \{ \hat{P}_i\}_n,
\end{equation}
where the last term in Equation \ref{eq:unique} includes all tuples of $\{ \hat{P}_i\}_n$. In other words, the gray line we plot in Figure  \ref{fig:pds_statevector} is the cardinality of $\{\hat{P}_i\}_\text{unique}$.

\subsection{Qubit Tapering}
\label{ssec:tapering}

In simple terms, qubit tapering seeks a decomposition of an $n$-qubit Hamiltonian $\hat{H}$ of the following form

\begin{equation}
    \hat{H} = \bigotimes_i^m \hat{H}_i'
\end{equation}
by exploring $\mathbb{Z}_2$ symmetries encountered in the Hamiltonian. If those are present, then there is a $\hat{H}_j'$, for $j\leq m$, that is completely defined within a subspace of the $2^n$-dimensional Hilbert space spanned by the eigenbasis of $\hat{H}$, while all other $\hat{H}_k'$, $k\neq j$, are limited to the space of a single qubit, and thus have eigenvalues $\pm 1$. The qubits associated with $k\neq j$ are thus tapered off, with one left to determine which combination of $\pm 1$ for the tapered qubits allows $\hat{H}_j'$ to contain the sought eigenvalue of the original Hamiltonian $\hat{H}$.

We used the XACC quantum computing framework\cite{xacc1, xacc2, xacc3} to carry out qubit tapering for the Hamiltonian associated to the singlet spin state, leading to a 5-qubit $\hat{H}'$ that houses the ground state energy. However, because XACC does not currently provide the associated tapered initial state, determining the basis state corresponding with tapered triplet initial state proved to be non-trivial within XACC, for which we turned to Pennylane.\cite{pennylane} A simple local modification to Pennylane to allow for initial states whose spatial orbitals are not doubly occupied, that is, a reference determinant that is not fully restricted, was effected to obtain the tapered initial state for the spin triplet.

\subsection{Quantum circuit parallel execution}

In turning to qubit tapering, we noticed that our the problem Hamiltonians in question could be defined by only 5 qubits. Given the architecture of the H1-1 system, one can measure 4 strings simultaneously in a single quantum circuit execution, each associated with a distinct QWC group. This means that the resulting counts refer now to 20-bit strings, where the corresponding basis states $|B_{20}\rangle $ are:

\begin{equation}
    |B_{20}\rangle = \bigotimes_{k=0}^{19} |b_k\rangle, b_k \in \{0, 1\}
\end{equation}

Let us define $P_1(k)$ as the probability distribution of finding $|b_k\rangle$ in each of the $k$-th qubit's basis state:

\begin{equation}
    P_1(k) = \begin{pmatrix}
        p_k^0 \\
        p_k^1
    \end{pmatrix},\quad p_k^0+p_k^1 = 1
\end{equation}

Thus, the probability distribution $P_{20}$ associated with 20-bit string outcomes 
can be written as

\begin{equation}
    P_{20} = \bigotimes_{k=0}^{19} P_1(k) = \bigotimes_{k=0}^{19} \begin{pmatrix}
        p_k^0 \\
        p_k^1
    \end{pmatrix}
\end{equation}

However, we are interested in the probabilities for the 5-bit strings that represent the QWC rotation operators, that is:

\begin{equation}
    \label{eq:probs}
    P_{20} = P_5(\mathbf{k}_1) \otimes P_5(\mathbf{k}_2)  \otimes P_5(\mathbf{k}_3)  \otimes P_5(\mathbf{k}_4) 
\end{equation}

Thus each probability distribution $P_5(\mathbf{k}_n)$, for $\mathbf{k}_n$ representing an index range over the qubits in the 5-qubit subspace associated with a QWC group in Equation \ref{eq:probs}, can be obtained as the marginal probabilities over the bits that are not in the desired bit string.

\section{Acknowledgements}

This work was supported by the “Embedding Quantum Computing into Many-body Frameworks for Strongly Correlated Molecular and Materials Systems” project, which is funded by the U.S. Department of Energy (DOE), Office of Science, Office of Basic Energy Sciences, the Division of Chemical Sciences, Geosciences, and Biosciences. This research used resources of the Oak Ridge Leadership Computing Facility at the Oak Ridge National Laboratory, which is supported by the Office of Science of the U.S. Department of Energy under Contract No. DE-AC05-00OR22725. This research used resources of the Compute and Data Environment for Science (CADES) at the Oak Ridge National Laboratory, which is supported by the Office of Science of the U.S. Department of Energy under Contract No. DE-AC05-00OR22725. 

\bibliographystyle{iopart-num}
\bibliography{main}

\end{document}


\title{Supplementary Information for ``Modeling singlet fission on a quantum computer''}

\author{Daniel Claudino,$^1$\footnote{\href{mailto:claudinodc@ornl.gov}{claudinodc@ornl.gov}} Bo Peng,$^{2}$\footnote{\href{mailto:peng398@pnnl.gov}{peng398@pnnl.gov}} Karol Kowalski$^{2}$ and Travis S. Humble$^{3}$}
\affiliation{
$^{1}$Computational Sciences and Engineering Division,\ Oak\ Ridge\ National\ Laboratory,\ Oak\ Ridge,\ TN,\ 37831,\ USA \\
$^{2}$Physical Sciences Division, Pacific Northwest National Laboratory, Richland, WA 99352, USA\\
$^{3}$Quantum Science Center,\ Oak\ Ridge\ National\ Laboratory,\ Oak\ Ridge,\ TN,\ 37831,\ USA}

\thanks{This manuscript has been authored by UT-Battelle, LLC, under Contract No.~DE-AC0500OR22725 with the U.S.~Department of Energy. The United States Government retains and the publisher, by accepting the article for publication, acknowledges that the United States Government retains a non-exclusive, paid-up, irrevocable, world-wide license to publish or reproduce the published form of this manuscript, or allow others to do so, for the United States Government purposes. The Department of Energy will provide public access to these results of federally sponsored research in accordance with the DOE Public Access Plan.}
\maketitle
\section{Electronic structure theory}

Here we discuss the most salient underpinnings of electronic structure theory relevant to this article. In general, quantum chemistry seeks stationary states that solve the non-relativistic Schr\"{o}dinger equation

\begin{equation}
    \label{eq:schrodinger}
    \hat{H}\Psi(\mathbf{x}_1, \cdots, \mathbf{x}_N; \mathbf{R}) = E\Psi(\mathbf{x}_1, \cdots, \mathbf{x}_N; \mathbf{R}), 
\end{equation}
where $\mathbf{x}_i$ are the spatial and spin coordinates of electron $i$ and $\mathbf{R}$ are the collective nuclear coordinates, while the Hamiltonian is given by

\begin{equation}
    \hat{H} = -\sum_i \frac{1}{2}\nabla_i^2 - \sum_{i, A}\frac{Z_A}{|\mathbf{r}_i-\mathbf{R}_A|} + \sum_{i,j}\frac{1}{|\mathbf{r}_i-\mathbf{r}_j|} + E_R,
\end{equation}
where indices $i$ and $A$ represent electrons and nuclei, respectively, and atomic units are adopted. The exclusively nuclear contribution, $E_R$, is simply the electrostatic interaction between pairs of nuclei, and the nuclear kinetic energy is neglected, assuming the validity of the Born-Oppenheimer approximation.

Alternatively, Equation \ref{eq:schrodinger} concerns with finding eigenstates of the molecular Hamiltonian in the absence of any external time-dependent field operator. As a common zero-th order approximation to the many-electron wave function, one chooses Hartree-Fock theory, which approximates the wave function by a Slater determinant $|\Phi \rangle$ built off of an antisymmetrized product of one-electron wave functions $\{\psi_k\}$, also known as spin-orbitals. This determinant has the property of being the lowest-energy single determinant for a given many-body wave function, filling the $N_e$-lowest  spin orbitals, in the case of $N_e$ electrons. Practically, these spin-orbitals are expanded in a basis of $N_b$ functions, hence yielding $N_v=N_b-N_e$ virtual (unoccupied) spin-orbitals. The ground state approximation afforded by the chosen basis can be obtained by an expansion over the Fock basis states characterized by the $N_e$ particle number:

\begin{equation}
    |\Psi_0 \rangle \approx  \left(1 + \sum_{ia}c_i^a a_a^\dagger a_i + \frac{1}{4}\sum_{ijab} c_{ij}^{ab}a_a^\dagger a_b^\dagger a_i a_j \cdots  \right) |\Phi \rangle,
\end{equation}
where $a^\dagger _p$ and $a_p$ are creation and annihilation operators associated with the $p$-th fermionic mode, with $\{i, j, k, \cdots\}$ indexing occupied spin-orbitals and $\{a, b, c, \cdots\}$ indexing virtual orbitals. Coupled cluster (CC)\cite{shavitt_bartlett_2009} theory provides an exponential parameterization of $| \Psi_0 \rangle $ in terms of the various excitation manifolds:

\begin{align}
    \label{eq:ops}
    |\Psi_0 \rangle &\approx e^{\hat{T}_1 + \hat{T}_2 + \cdots} |\Phi\rangle, \nonumber \\
    \hat{T}_1 &= \sum_{ia} t_i^a \hat{a}_a^\dagger \hat{a}_i,  \nonumber \\
    \hat{T}_2 &=  \frac{1}{4}\sum_{ijab} t_{ij}^{ab} \hat{a}_a^\dagger \hat{a}_b^\dagger \hat{a}_i \hat{a}_j, \nonumber \\
    & \vdots \nonumber \\
    \hat{T}_n &= \left(\frac{1}{n!}\right)^2\sum_{ij\cdots p, ab \cdots q} t_{ij\cdots p}^{ab\cdots q} \hat{a}_a^\dagger \hat{a}_b^\dagger \cdots \hat{a}_q^\dagger \hat{a}_i \hat{a}_j \cdots \hat{a}_p,
\end{align}
where the amplitudes $t$ are scalars. The exact ground state for the chosen basis includes cluster operators $\hat{T}_n$ with $n$ up to $N_e$. One can target the excited state $| \Psi_k \rangle $ via the excitation operator $\hat{R}_k$ by the equation-of-motion (EOM)\cite{eom} variant of CC:

\begin{equation}
    \label{eq:r}
    | \Psi_k \rangle = \hat{R}_k | \Psi_0 \rangle
\end{equation}
with the corresponding excitation energy $\Delta E_k = E_k-E_0$ given by

\begin{equation}
    \left[\hat{H}, \hat{R}_k \right ] | \Psi_0 \rangle = \Delta E_k  \hat{R}_k| \Psi_0 \rangle
\end{equation}
where $\hat{R}_k$ is expanded in a basis of excitation operators, similarly to $\hat{T}_n$ in Equation \ref{eq:ops}. Thus, in the present context, (EOM)CCSD, (EOM)CCSDT, and (EOM)CCSDTQ mean $\hat{T}$ in Equation \ref{eq:ops} (and $\hat{R}_k$, Equation \ref{eq:r}) include up to double ($\hat{T}_2$), triple ($\hat{T}_3$), and quadruple ($\hat{T}_4$) excitation operators, respectively. Because the H$_4$ molecule has four electrons, the operators $\hat{T}$ and $\hat{R}$ in (EOM)CCSDTQ span the entire subspace for the desired particle number, thus recovering the exact eigenspectrum in the one-particle basis of choice.

\section{State energies}

Table \ref{tab:quantum} reports the energy values obtained from quantum computations with a noiseless statevector simulator and the H1-1 system and Table \ref{tab:cc} shows the energies computed classically from various levels of CC and EOM-CC theories.

\begin{table}[h]
    \centering
    \caption{Energies (in a.u.) for the $S_0$, $S_1$, and $T_0$ states computed with a noiseless statevector simulator and the H-1 quantum hardware.}
\begin{tabular}{c c c c}
\toprule
 &  $S_0$  & $S_1$ & $T_0$ \\
 \hline
    Noiseless   & -1.897780   & -1.856543   & -1.881876   \\
    H1-1        & -1.898401   & -1.864233   & -1.881865   \\
 \hline
\end{tabular}
    \label{tab:quantum}
\end{table}

\begin{table}[h]
    \centering
    \caption{Coupled cluster energies (in a.u.) for the $S_0$ and $T_0$ states and the EOM coupled cluster energies (in eV) for the $S_0\rightarrow S_1$ transition.}
\begin{tabular}{c c c c }
\toprule
 &  $S_0$  &  $T_0$ & $S_0 \rightarrow S_1$ \\
 \hline
    (EOM)CCSDT  &  -1.916086   &  -1.881875  &  0.13139                \\
    (EOM)CCSDT  &  -1.908069   &  -1.881876  &  0.06631                \\
    (EOM)CCSDTQ &  -1.897781   &  -1.881876  &  0.04117                 \\
 \hline
\end{tabular}
    \label{tab:cc}
\end{table}

Table \ref{tab:determinants} provides the contribution of each doubly-excited determinant in the $S_1$ state for the different levels of EOM-CC theory.

\begin{table}[h]
    \centering
    \caption{Contributions from doubly-excited determinants for the different levels of EOM-CC theory.}
\begin{tabular}{c c c c }
\toprule
 &  CCSD  &  CCSDT &  CCSDTQ\\
 \hline
    $|\Phi_{12}^{34}\rangle$                & 0.24438 & 0.23674 & 0.14833  \\
    $|\Phi_{2\bar{2}}^{3\bar{3}}\rangle$    & 0.57627 & 0.51145 & 0.29432 \\
    $|\Phi_{1\bar{1}}^{3\bar{3}}\rangle$    & 0.28854 & 0.32796 & 0.25263 \\
    $|\Phi_{2\bar{2}}^{4\bar{4}}\rangle$    & 0.27117 & 0.30838 & 0.23670 \\
    $|\Phi_{1\bar{1}}^{4\bar{4}}\rangle$    & 0.36685 & 0.34619 & 0.18536 \\
    $|\Phi_{1\bar{2}}^{3\bar{4}}\rangle$    & 0.20310 & 0.14869 &     ---      \\
 \hline
\end{tabular}
    \label{tab:determinants}
\end{table}

Intuitively, the $|\Phi_{1\bar{1}}^{4\bar{4}}\rangle$ deteminant should be the least important contribution among the determinants that are a pair excitation away from the reference (HF) determinant. However, EOM-CCSD and EOM-CCSDT both place an undue importance on this determinant, which in these two instances provides the second largest contribution, second only to the determinant that represents the HOMO-LUMO transition ($|\Phi_{2\bar{2}}^{3\bar{3}}\rangle$). This inadequacy is remedied by EOM-CCSDTQ, which also does not find any support in the determinant of the type $|\Phi_{1\bar{2}}^{3\bar{4}}\rangle$, whose contribution is non-negligible in both EOM-CCSD and EOM-CCSDT.

\section{Error analysis}

Both the state preparation stage as well as the change of basis to the QWC rotation operators are carried out solely by single-qubit gates, whose error rates are $1 \times 10^{-5} - 3 \times 10^{-4}$. However, the state preparation and measurement (SPAM) error is 1-2 orders of magnitude higher than those of one-qubit gates ($2 \times 10^{-3} - 5 \times 10^{-3}$), hence they are expected to dominate the error observed from quantum channels. A simple, but effective, classically inspired correction is to assume there is a matrix $\mathbf{A}$ that maps ideal probability distributions $P_\text{ideal}$ onto the observed, noisy ones $P_\text{noisy}$

\begin{equation}
    P_\text{noisy} = \mathbf{A}P_\text{ideal} \Rightarrow P_\text{ideal} = \mathbf{A}^{-1} P_\text{noisy}.
\end{equation}

Each element $A_{ij}$ corresponds to the probability of preparing the bit string indexed by $i$ but to measure the $j$-th string. Since we use all 20 qubits available in the H1-1 system (except for a single experiment for both spin states), building the entire $\mathbf{A}$ is not desirable, neither from a quantum viewpoint, since it takes $2^{20}$ distinct experiments, nor classically, as it involves storage of $2^{20} \times  2^{20} = 2^{40}$ double precision elements, or $2^{19}(2^{20+1})$ unique elements, which demands terabytes of storage. Fortunately, we need only the relevant elements in $\mathbf{A}$, which we can access as shown in Ref. \citenum{scalable_mitigation}. In that approach, $\mathbf{A} = \bigotimes_{k=0}^{19} \mathbf{S}^k $, where $\mathbf{S}^k$ stores the probabilities of each outcome given the initial state of the $k$-th qubit:

\begin{equation}
    \mathbf{S}^k = \begin{pmatrix}
        p^k_{00} & p^k_{01} \\
        p^k_{10} & p^k_{11}
    \end{pmatrix}
\end{equation}
where $p^k_{ij}$ indicates the probability that the qubit in state $|i\rangle$ be measured in the state $|j\rangle$, $i, j \in \{ 0, 1\}$.

In an ideal setting, all trapped ions are equivalent, and so are the corresponding qubits, which makes all $\mathbf{S}_i$ numerically identical. While the ions are equivalent, the vicinity around each one is not, which can impact each qubit slightly differently. However, Quantinuum reports the minimum, maximum, and average SPAM errors over all qubits. Furthermore, ideally we would also observe $p^k_{00}=p^k_{11}$ and $p^k_{01}=p^k_{10}$, and experiments we conducted did not reveal any noticeable bias, so we take these equalities as valid. Therefore, assumning the equivalence of the qubits at hand, if the probability of flipping any given qubit is $p$, then

\begin{equation}
    \mathbf{S}^k = \begin{pmatrix}
        1-p & p \\
        p & 1-p
    \end{pmatrix}
\end{equation}

Although Ref. \citenum{scalable_mitigation} also details how to include effects of correlated noise, we assume it is non-existing/negligible, as Ref. \citenum{h1_sheet} affirms absence of crosstalk errors. First, we recognize that $\mathbf{A}^{-1} = \bigotimes_{i=0}^{19} \left( \mathbf{S}^k\right)^{-1} $, with:

\begin{equation}
    (S_{ij}^k)^{-1} = \frac{p-\delta_{ij}}{2p-1}
\end{equation}

Thus, the error-mitigated probability $\tilde{P}_i$ of obtaining the $k$-th bit string is:

\begin{equation}
    \label{eq:final_prob}
    \tilde{P}_i = \sum_{j \in B} \prod_{k=0}^{19} (S^k_{ij})^{-1} P_j
\end{equation}
where $B$ is the set of all bit strings indices. Because the probability distribution is highly sparse, we take $B = \{j \in [0, 2^{20}] : P_j \neq 0\}$. Finally, one needs to ensure the mitigated probability distributions are normalized. 

However, we can see in Table \ref{tab:mitigation} that mitigating SPAM errors ultimately imparts little effect to the final energies.

\begin{table}[h]
    \centering
    \caption{Energies (in a.u.) for the state $S_0$, $S_1$, and $T_0$ computed with probabilities mitigated according to Equation \ref{eq:final_prob} for different bit-flip probabilities $p$.}
\begin{tabular}{c c c c}
\toprule
 $p$ &  $S_0$  & $S_1$ & $T_0$ \\
 \hline
    0                   &   -1.898402     &   -1.864233     &   -1.881865 \\
    $1\times 10^{3}$    &   -1.898403     &   -1.864241     &   -1.881869 \\
    $5\times 10^{3}$    &   -1.898407     &   -1.864278     &   -1.881874 \\
 \hline
\end{tabular}
    \label{tab:mitigation}
\end{table}

Turning to the qsim\cite{qsim} simulator and considering only sampling noise, we run simulations measuring a single QWC rotation string at a time (serial) and four QWC rotation strings (parallel), as discussed in the main text, and we observe the energies reported in Table \ref{tab:sampling}.

\begin{table}[h]
    \centering
    \caption{Energies (a.u.) for the state $S_0$, $S_1$, and $T_0$ computed from the qsim simulator with sampling noise (shots).}
\begin{tabular}{c c c c}
\toprule
  &  $S_0$  & $S_1$ & $T_0$ \\
 \hline
    Noiseless               &   -1.897780   &   -1.856543    &   -1.881876   \\
    Serial (8192 shots)     &   -1.897768   &   -1.854326    &   -1.881877 \\
    Serial (10$^5$ shots)   &   -1.897779   &   -1.855315    &   -1.881875 \\
    Parallel (8192 shots)   &   -1.899059   &   -1.867473    &   -1.881857 \\
    Parallel (10$^5$ shots) &   -1.899086   &   -1.868069    &   -1.881861 \\
 \hline
\end{tabular}
    \label{tab:sampling}
\end{table}

As we assert in the main text, PDS($K$) is a theory primarily concerned with the ground state, corroborated by the much slower convergence of the $S_1$ state. Table \ref{tab:sampling} provides evidence that the bound to the energy of this state is much more vulnerable to sampling error. The absence of two-qubit gate errors, the one-qubit gate errors being negligible in the scale of the other noise sources, and mitigating SPAM errors having a meager influence, Table \ref{tab:sampling} leads us to the conclusion that sampling noise is the primary culprit for the deviation of the estimate for the $S_1$ state from the reference values, and consequently for the $S_0\rightarrow S_1$ transition.

\includegraphics[scale=0]{figures/levels.pdf}
